\documentclass{article}
\usepackage{graphicx} 
\usepackage{amssymb}
\usepackage{amsmath}
\usepackage{biblatex}
\usepackage{arxiv}

\usepackage{authblk}
\usepackage{hyperref}
\usepackage{fancyhdr}
\title{Quantum Signal Processing Phase Factorization via Logarithmic Derivatives}
\author{Christopher Peng  \\ \texttt{christopherbpeng@gmail.com}}
\affil{Independent Researcher}
\date{October 2026}

\begin{document}

\maketitle

\begin{abstract}
Quantum signal processing (QSP) requires the efficient computation of phase factors associated with a polynomial transformation. One existing stable factorization method avoids direct root finding but relies on Prony's method to recover a polynomial from weighted power sums of its roots~\cite{ying2022stable}. We present a modification of this approach that eliminates Prony's method by using the logarithmic derivative of an associated polynomial. Because the logarithmic derivative has the same residue at every root after accounting for multiplicity, its Fourier coefficients directly yield the unweighted power sums of the roots inside the unit disk. The Newton--Girard identities then recover the coefficients of the corresponding polynomial without computing its roots. The resulting algorithm has $O(d\log^2 d)$ computational complexity and simplifies the factorization procedure of the original method. In numerical experiments on Hamiltonian simulation, eigenstate filtering, matrix inversion, and the Fermi--Dirac operator, our implementation computes phase factors in seconds with maximum errors on the order of $10^{-12}$.
\end{abstract}

\section{Introduction}

\subsection{Background}

Quantum signal processing (QSP) is a framework for implementing polynomial transformations of non-unitary operators with quantum circuits.

Let $A$ be a $2^n \times 2^n$ Hermitian matrix with $\|A\|_2 < 1$. We assume $A$ is Hermitian for simplicity; more general cases are covered in~\cite{dong2021efficient, gilyen2019quantum}. To work with $A$ on a quantum computer, we encode it in the upper-left block of a larger Hermitian unitary matrix
\[
U_A \equiv \begin{bmatrix} A & * \\ * & * \end{bmatrix}
\]
of size $2^{m+n} \times 2^{m+n}$, called a block encoding of $A$.

We often want to implement a transformation $f(A)$ for some function $f$ defined on $[-1,1]$ with $\|f\|_\infty < 1$. To do so with a quantum circuit, we must construct a block encoding $U_{f(A)}$ of $f(A)$ from $U_A$.

First, the even and odd parts of $f(x)$ are approximated by polynomials $a^e(x)$ and $a^o(x)$, which are later combined using a linear combination of unitaries (LCU)~\cite{childs2017quantum}.

It remains to compute the phase factors for each polynomial of definite parity. The quantum signal processing theorem~\cite{low2017optimal, gilyen2019quantum} guarantees that for every real polynomial $a(x)$ of degree $d$ with parity $d \pmod{2}$ and $\|a\|_\infty < 1$, there exists a sequence of parameters $\Phi = (\phi_0, \dots, \phi_d) \in [-\pi, \pi]^{d+1}$, called phase factors, such that $a(x) = \Re(p(x))$, where
\[
U(x, \Phi) = \begin{pmatrix} p(x) & r(x) \\ -r^*(x) & p^*(x) \end{pmatrix} = \left(\prod_{k=0}^{d-1} e^{i\phi_kZ} e^{i \arccos(x)X}  \right)e^{i\phi_dZ}
\]
and
\[
X := \begin{pmatrix} 0 & 1 \\ 1 & 0 \end{pmatrix}, \quad Z := \begin{pmatrix} 1 & 0 \\ 0 & -1 \end{pmatrix}
\]
are the Pauli matrices.

\subsection{Notation}
\label{sec:notation}

As in~\cite{ying2022stable}, we let $a(x)$ be the target polynomial of degree $d$ and use the notational convention
\[
a(x) = \Re(p(x)), \quad c(x) = \Im(p(x)), \quad b(x) = \Re(r(x)), \quad d(x) = \Im(r(x)).
\]
For a function $f$ on $[-1,1]$, we also use the lifted functions
\[
f(t) := f(\cos t), \qquad f(z) := f\left(\frac{z+z^{-1}}{2}\right),
\]
so that $f(z)$ agrees with $f(t)$ at $z = e^{it}$.

\subsection{Previous Work}
\label{sec:prev-work}

One approach to computing phase factors is polynomial factorization~\cite{gilyen2019quantum}, which traditionally relies on numerically unstable root finding. However, root finding can be avoided by applying Prony's method to weighted power sums of the roots, as in the following algorithm of~\cite{ying2022stable}:

\begin{enumerate}
\item Construct a polynomial approximation $a(t)$ using the DFT and truncation.
\item Choose $b(t)$ of the form $b_d\sin(dt) + b_{d-2}\sin((d-2)t) + \dots$ with suitable high-frequency terms.
\item Let $g(z) = \frac{1}{1-a^2(z) - b^2(z)}$, the reciprocal of a degree-$4d$ Laurent polynomial, with poles $\{\xi_j\}$.
\item Compute the Fourier coefficients of $g(t)$ by numerical quadrature with the DFT.
\item Apply Prony's method to these Fourier coefficients to construct the characteristic polynomial $e(z) = z^{-d}\prod_{|\xi_j|<1} (z - \xi_j)$. We often work with the polynomial $m(z) = z^de(z)$ as well.
\item Use $e(z)$ to construct $p(z)$ and $r(z)$.
\item Extract the phase factors using a recursive algorithm.
\end{enumerate}

Thus, the phase factors are constructed from $a(x)$ and $b(x)$.

Beyond factorization~\cite{gilyen2019quantum,  Haah2019, ying2022stable} and iterative methods~\cite{dong2021efficient, DongLinNiEtAl2024_iqsp,DongLinNiEtAl2024_newton, qspbasics}, recent work has connected QSP to nonlinear Fourier analysis~\cite{alexis2024quantum, alexis2024infinite, ni2024fast}; in particular, \cite{ni2025inverse} computes phase factors in $O(d\log^2 d)$ time via divide-and-conquer layer stripping. Berntson and S\"{u}nderhauf \cite{berntson2025complementary} derive a contour integral representation of a canonical complementary polynomial and use it to build an FFT-based algorithm with explicit error guarantees. We refer to~\cite{lin2025mathematical} for a recent survey.

\subsection{Contribution}

The main contribution of this paper is a modification of the stable factorization algorithm of~\cite{ying2022stable} that eliminates the need for Prony's method, accelerating the construction of the characteristic polynomial in step 5 of Section~\ref{sec:prev-work}. The remaining steps follow the pipeline of~\cite{ying2022stable}, accelerated using the divide-and-conquer phase stripping algorithm of~\cite{ni2025inverse}.

Ying's approach extracts weighted power sums of the roots of the characteristic polynomial from Fourier coefficients of the reciprocal $1/(1-a^2-b^2)$. We observe that the Fourier coefficients of a logarithmic derivative instead give unweighted power sums, so the Newton--Girard identities recover the polynomial directly. This factorization coincides with that of~\cite{berntson2025complementary}.

The resulting algorithm remains easy to implement, achieves accuracy on the order of $10^{-12}$, and has $O(d\log^2 d)$ complexity.

The rest of the paper is organized as follows. Section~\ref{sec:Newton-identities} reviews the Newton--Girard identities, Section~\ref{sec:algorithm} describes the algorithm, Section~\ref{sec:results} presents the numerical results, and Section~\ref{sec:conclusion} offers concluding remarks.

\section{Review of Newton--Girard Identities}
\label{sec:Newton-identities}

Let
\[
m(z)=z^n+c_1z^{n-1}+\cdots+c_n
=\prod_{j=1}^n(z-\xi_j)
\]
be a monic polynomial with roots $\xi_1,\ldots,\xi_n$, and define the power sums
\[
S_k=\sum_{j=1}^n\xi_j^k.
\]
The Newton--Girard identities relate the power sums $S_k$ to the coefficients of $m$. Setting $c_0=1$, they state that for $1\leq k\leq n$,
\[
k c_k+\sum_{j=1}^k c_{k-j}S_j=0,
\]
or equivalently,
\[
c_k=-\frac{1}{k}\sum_{j=1}^k c_{k-j}S_j.
\]

Thus, given the first $n$ power sums $S_1,\ldots,S_n$, the coefficients of $m$ can be recovered recursively. In our setting, $n=2d$, and the Fourier coefficients of $l$ provide precisely these power sums. Consequently, the coefficients of the characteristic polynomial $m(z)$, and hence those of $e(z)=z^{-d}m(z)$, can be computed without finding any of its roots.

Computing each $c_k$ with this recursion requires $O(k)$ operations, so recovering all $n$ coefficients requires
\[
\sum_{k=1}^n O(k)=O(n^2)=O(d^2)
\]
operations. Section~\ref{sec:fast-newton} reduces this cost to $O(d\log d)$.

\section{Algorithm}
\label{sec:algorithm}

\subsection{Key Components}

The choice $b\neq0$ in Ying's algorithm is primarily motivated by the need to ensure that the Prony matrix has rank $2d$. Since our approach does not form a Prony matrix, this requirement is unnecessary. We therefore take $b(t)=0$, which reduces the factorization problem to the construction of a single complementary polynomial.

\subsection{Polynomial Approximation}

First, we scale $f$ to have $\|f\|_\infty = 0.3$ and construct an accurate polynomial approximation $a(x)$. Following~\cite{ying2022stable}, we apply the FFT to samples of $f(t)$, truncate at a sufficiently high degree, and transform back to obtain a trigonometric polynomial.

Specifically, we sample $f$ on an equally spaced grid $\{t_n\}$ of size $N$ and apply the FFT to $\{f(t_n)\}$. We choose the frequency $d$ such that all Fourier coefficients with $|k| > d$ are smaller than a threshold times the largest coefficient in absolute value, and we adjust $d$ to have the same parity as $f(x)$. As in~\cite{ying2022stable}, we set this threshold to about $10^{-12}$, which matches the accuracy of the QSP algorithm. Setting the Fourier coefficients with $|k| > d$ to zero yields the coefficients of the approximating polynomial.

\subsection{Characteristic Polynomial}

Let $a \in \mathbb{R}[x]$ have degree $d$, parity $d \pmod{2}$, and $\|a\|_\infty < 1$. Recall from Section~\ref{sec:notation} that $a(z)$ denotes the Laurent lift of $a(x)$, and define
\[
h(z) = z^{2d}\left(1-a(z)^2\right).
\]
Since $a(z)$ is a symmetric Laurent polynomial with terms from $z^{-d}$ to $z^{d}$, $h$ is a palindromic polynomial of degree $4d$ with nonzero constant and leading coefficients. Let $\{\xi_j\}$ be the roots of $h$, as in Section~\ref{sec:prev-work}.

The roots of $h$ have the following structure; see also~\cite{qspbasics}. Since $a(z) = a(z^{-1})$, the roots occur in reciprocal pairs $\xi_j$, $1/\xi_j$. Because $a(e^{it}) = a(\cos t)$ is real and $\|a\|_\infty<1$, we have $1-a(z)^2>0$ for all $|z| = 1$, so $h$ has no roots on the unit circle. Finally, the nonzero constant term rules out a root at $0$, and the nonzero leading coefficient ensures that $h$ has exactly $4d$ roots. Hence exactly $2d$ roots lie inside the unit disk and $2d$ lie outside.

The goal of the factorization process is to determine the coefficients of $e(z)$, defined in Section~\ref{sec:prev-work}, without computing its roots.

Define
\[
l(z) = \frac{h'(z)}{h(z)} = \frac{\mathrm{d}}{\mathrm{d}z} \log h(z).
\]
Its poles are exactly the roots of $h$. Writing $h(z) = C\prod_{j} (z - \xi_j)$, we have
\[
l(z) = \sum_{j} \frac{1}{z - \xi_j}.
\]

Let $\hat{l}_k = \frac{1}{2\pi}\int_0^{2\pi} l(e^{it}) e^{-ikt}\,\mathrm{d}t$ denote the $k$th Fourier coefficient of $l(e^{it})$. Following~\cite{ying2022stable}, for $k \le -1$ we have
\begin{align}
    \hat{l}_k &=\frac{1}{2\pi i} \oint _{|z| = 1}\frac{l(z)}{z^{k+1}}\,\mathrm{d}z \\
    &= \frac{1}{2\pi i} \oint_{|z| = 1}\sum_{j} \frac{1}{z - \xi_j}z^{-(k+1)} \,\mathrm{d}z \\
    &= \frac{1}{2\pi i} \oint_{|z| = 1}\sum_{|\xi_j| < 1} \frac{1}{z - \xi_j}z^{-(k+1)} \,\mathrm{d}z \\
    &= \sum_{|\xi_j| < 1} \xi_j^{-(k+1)} \\
    &=: S_{-(k+1)},
\end{align}
where the contour integrals are taken counterclockwise over the unit circle. The third equality holds because, for $k \le -1$, $z^{-(k+1)}$ is a polynomial, so the terms with $|\xi_j| > 1$ are analytic in the unit disk and integrate to zero; the fourth follows from the residue theorem. Therefore
\[
\left(\hat{l}_{-2}, \hat{l}_{-3}, \dots, \hat{l}_{-2d-1}\right) = \left(S_1, S_2, \dots, S_{2d}\right).
\]

Define
\[
m(z) := z^de(z) = \prod_{|\xi_j| < 1} (z - \xi_j).
\]
Then $m$ is the monic degree-$2d$ polynomial whose roots are precisely the roots of $h$ inside the unit disk, and the $S_k$ are the first $2d$ power sums of the roots of $m$. The Newton--Girard identities then recover the coefficients of $m$, and hence those of $e$.

In Ying's algorithm, the Fourier coefficients of $1/(1-a^2-b^2)$ give the same power sums weighted by unknown residues, which is why Prony's method is needed. The logarithmic derivative has residue $1$ at every simple root, so the weights disappear and the Newton--Girard identities apply directly.
\subsection{Fast Newton--Girard Recovery}
\label{sec:fast-newton}

The recursion of Section~\ref{sec:Newton-identities} costs $O(d^2)$ operations. We reduce this to $O(d\log d)$ using fast power series arithmetic.

Recall that $m(z) = z^{2d} + c_1 z^{2d-1} + \cdots + c_{2d} = \prod_{j}(z-\xi_j)$, where the product is over the $2d$ roots of $h$ inside the unit disk. Reversing the order of the coefficients gives
\[
\tilde m(x) := x^{2d}\, m(1/x) = \prod_{j} (1 - \xi_j x) = \sum_{k=0}^{2d} c_k x^k,
\]
with $c_0 = 1$. Taking logarithms and using $\log(1-\xi x) = -\sum_{k\ge1} \xi^k x^k/k$ for $|x|$ small, we obtain
\[
\log \tilde m(x) = -\sum_{j}\sum_{k\ge1}\frac{\xi_j^k}{k}x^k = -\sum_{k\ge1}\frac{S_k}{k}x^k,
\qquad\text{so}\qquad
\tilde m(x) = \exp\Bigl(-\sum_{k\ge1}\frac{S_k}{k}x^k\Bigr).
\]
Since $\tilde m$ has degree $2d$, its coefficients $c_0,\dots,c_{2d}$ are determined by $S_1,\dots,S_{2d}$: they are the first $2d+1$ Taylor coefficients of the right-hand side. (Differentiating $\log\tilde m$ and comparing coefficients recovers exactly the Newton--Girard recursion of Section~\ref{sec:Newton-identities}.) We compute this exponential modulo $x^{2d+1}$ by Newton iteration with FFT-based polynomial multiplication, which costs $O(d\log d)$ operations~\cite{brent1978fast}. Reading the coefficients in reverse order then gives $m(z) = \sum_{k=0}^{2d} c_k z^{2d-k}$.

Note that on the unit circle, $\log(1 - a^2) = \log \alpha - \sum_{k \ge 1} (S_k/k)(z^k + z^{-k})$, where $\alpha$ is the normalization constant defined in Section~\ref{p_recovery}. Thus the power sums obtained from $h'/h$ are, up to the factor $-k$, the Fourier coefficients of $\log(1 - a^2)$. The construction above therefore produces the same canonical complementary polynomial as the FFT-based algorithm of Berntson and Sünderhauf~\cite{berntson2025complementary}. The two differ in how it is computed: we sample the rational function $h'/h$ rather than a logarithm, and we exponentiate in coefficient space modulo $x^{2d+1}$ rather than pointwise on a grid. The result is a drop-in replacement for the Prony step of \cite{ying2022stable}.

\subsection{Recovering $p(z)$ from $e(z)$}
\label{p_recovery}

Once $m(z)$, and hence $e(z)$, is known, $c(z)$ and $d(z)$ are given by~\cite{ying2022stable}
\[
c(z) = \sqrt{\alpha} \cdot \frac{e(z) + e(1/z)}{2}, \qquad d(z) = \sqrt{\alpha} \cdot \frac{e(z) - e(1/z)}{2i},
\]
where the normalization constant
\[
\alpha = \frac{1-a^2(z)}{e(z)e(1/z)}
\]
is independent of $z$; we evaluate it at $z=1$. Since $b(z)=0$, we define $p(z) = a(z) + ic(z)$ and $r(z) = id(z)$.

\subsection{Phase Factors from $p(z)$}

The phase factors can then be extracted as in ~\cite{gilyen2019quantum}, written in terms of $t$.

Let $p_n(t)$ and $r_n(t)$ be trigonometric polynomials of degree $n$ with degree-$n$ coefficients $P_n$ and $R_n$, respectively. The phase factor $\phi_n$ is determined by $e^{2i\phi_n} = P_n /R_n$, and we compute
\[
\begin{pmatrix} p_{n-1}(t) & r_{n-1}(t) \end{pmatrix} =\begin{pmatrix} p_{n}(t) & r_{n}(t) \end{pmatrix} \begin{pmatrix} e^{-i\phi_n} & 0 \\ 0 & e^{i\phi_n} \end{pmatrix} \begin{pmatrix} \cos t & -i\sin t \\ -i\sin t & \cos t \end{pmatrix}.
\]
This choice of $\phi_n$ cancels the top coefficients, so $p_{n-1}$ and $r_{n-1}$ have degree $n-1$. Starting with $p_d(t) = p(t)$ and $r_d(t) = r(t)$ and iterating from $n=d$ down to $1$ yields all the phase factors except for $\phi_0$. Since the remaining pair is $p_0 = e^{i\phi_0}$, $r_0=0$, we set $\phi_0 = \arg p_0$.

This layer-peeling recursion can be carried out in $O(d\log^2d)$ time with a divide-and-conquer strategy, as in~\cite{ni2025inverse}. Since each step only couples coefficient $j$ with coefficient $j+2$, the leading coefficient after $k$ steps depends only on the top $2k+1$ input coefficients. We therefore run the first half of the iterations on the top half of the coefficients, apply the combined $2\times 2$ transfer matrix to the full-length arrays by FFT convolution, and recurse on the half-length remainder.

\subsection{Implementation}
\label{sec:implementation}

In our implementation, we scale so $\|a\|_\infty = 0.3$, following the numerical stability considerations of~\cite{ying2022stable}. This keeps $h$ bounded away from $0$ on the unit circle, so that $l$ remains well-behaved on the integration contour. We use about $40d$ samples of $l(z)$ to compute its Fourier coefficients accurately.

We evaluate the error of the resulting QSP circuit using the reference implementation in QSPPACK~\cite{dong2021efficient}.

\section{Numerical Results}
\label{sec:results}

\subsection{Setup}

We implemented the proposed factorization algorithm in standard double-precision arithmetic. All numerical experiments were performed on a laptop with a 2.00~GHz AMD Ryzen AI 7 350 processor. The implementation uses FFTs to compute the Fourier coefficients of the logarithmic derivative and then applies the Newton--Girard identities to recover the characteristic polynomial.

The experiments show that the proposed method remains numerically stable at polynomial degrees exceeding $10^4$ while requiring only a few seconds of computation on a standard laptop. The errors remain small throughout all tested ranges, and the observed growth in runtime is consistent with the predicted near-linear $O(d\log^2d)$ complexity. Recall that the target polynomial is scaled to have $\|a\|_\infty=0.3$, as described in Section~\ref{sec:implementation}.

The code used for these experiments is available at \url{https://github.com/chessbear18/QSP-Phase-Factorization-via-Logarithmic-Derivatives}.

\subsection{Examples}

We evaluate the proposed algorithm on four applications considered in~\cite{ying2022stable}: Hamiltonian simulation, eigenstate filtering, matrix inversion, and the Fermi--Dirac operator. For each application, we vary the parameter controlling the approximation difficulty and record the degree $d$ of the polynomial approximation, the total phase factor construction time, and the phase factor error. The phase factor error is the relative $L_\infty$ error between the target polynomial and the polynomial reconstructed from the computed phase factors, evaluated at 20 random points on the sampling grid plus the maximizer of $|a|$. Since only 21 points are used, the measured maximum is a noisy estimate of the true maximum error.

In all cases, the polynomial approximation step takes under 3 seconds, and the function approximation error is below $4 \times 10^{-11}$. Following~\cite{ying2022stable}, both are excluded from the reported times and errors. We use the threshold $\epsilon = 10^{-12}$ in all applications.

\subsubsection{Hamiltonian Simulation}

Consider a Hamiltonian $H$ with $\|H\|_2\leq 1$. Hamiltonian simulation for time $\tau$ requires approximating
\[
f(x)=e^{-i\tau x},
\]
whose real and imaginary parts are
\[
\Re(f(x))=\cos(\tau x),\qquad
\Im(f(x))=-\sin(\tau x).
\]

For simplicity, we target $\cos(\tau x)$ and $\sin(\tau x)$. We perform experiments with $\tau=1000,2000,3000,4000,5000$.

\begin{figure}[!h]
\centering
\includegraphics[width=0.8\linewidth]{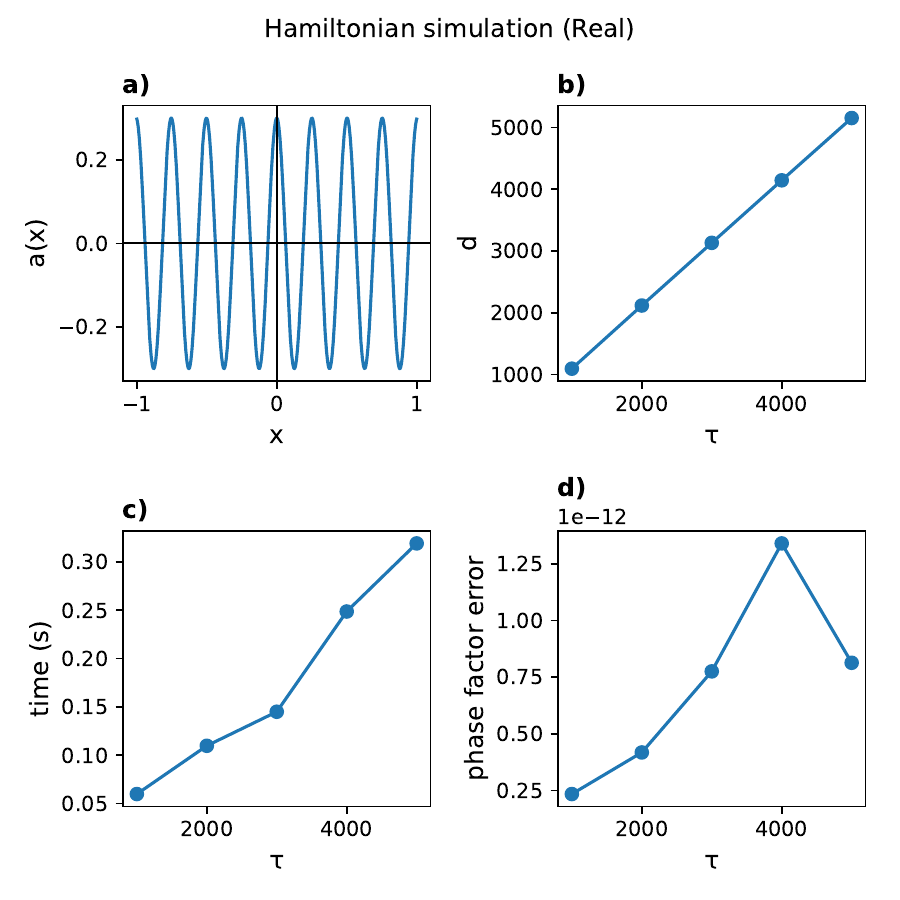}
\caption{Hamiltonian simulation, real part. (a) $a_{\mathrm{re}}(x)$ for $\tau=25$. (b) The degree $d$ of the polynomial approximation as a function of $\tau$. (c) The total phase factor construction time in seconds as a function of $\tau$. (d) The relative $L_\infty$ error of the phase factor reconstruction as a function of $\tau$.}
\label{ham_sim}
\end{figure}

\begin{figure}[!h]
\centering
\includegraphics[width=0.8\linewidth]{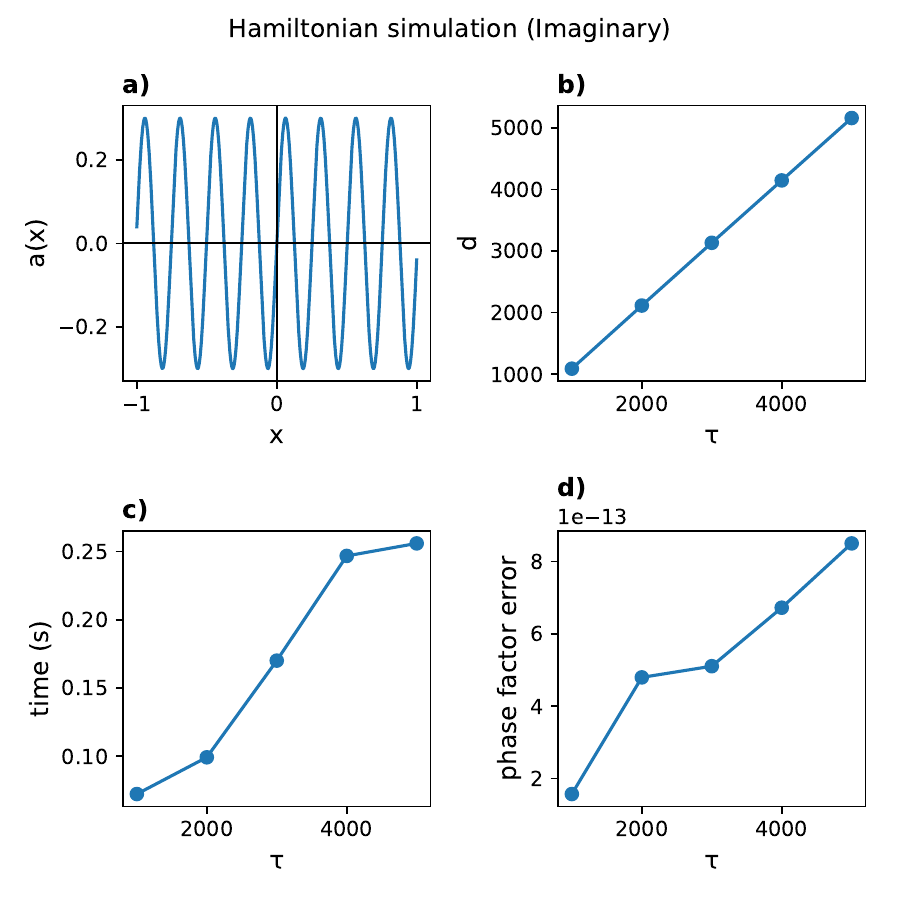}
\caption{Hamiltonian simulation, imaginary part. (a) $a_{\mathrm{im}}(x)$ for $\tau=25$. (b) The degree $d$ of the polynomial approximation as a function of $\tau$. (c) The total phase factor construction time in seconds as a function of $\tau$. (d) The relative $L_\infty$ error of the phase factor reconstruction as a function of $\tau$.}
\label{ham_sim_im}
\end{figure}

In both the real and imaginary cases, the degree grows linearly in $\tau$ with slope about $1$ (Figures~\ref{ham_sim} and~\ref{ham_sim_im}). By the Jacobi--Anger expansion, the Chebyshev coefficients are the Bessel functions $2J_k(\tau)$, which remain large up to $k \approx \tau$ and then decay rapidly. Our results are consistent with the standard bound $d = \tau + O(\tau^{1/3} \log (1/\epsilon)^{2/3})$~\cite{low2017optimal}.

\subsubsection{Eigenstate Filtering}

For a fixed gap $\Delta$, consider the filtering function centered at the origin
\[
f(x)=
\frac{
T_k\left(-1+2\frac{x^2-\Delta^2}{1-\Delta^2}\right)
}{
T_k\left(-1+2\frac{-\Delta^2}{1-\Delta^2}\right)
},
\]
where $T_k$ denotes the $k$th Chebyshev polynomial. As in~\cite{ying2022stable}, we set $k=20/\Delta$ so that $f(x)$ is negligible outside a $\Delta$-neighborhood of the origin.

We perform experiments with $1/\Delta=12.5,25,50,100,200$.

\begin{figure}[!h]
\centering
\includegraphics[width=0.8\linewidth]{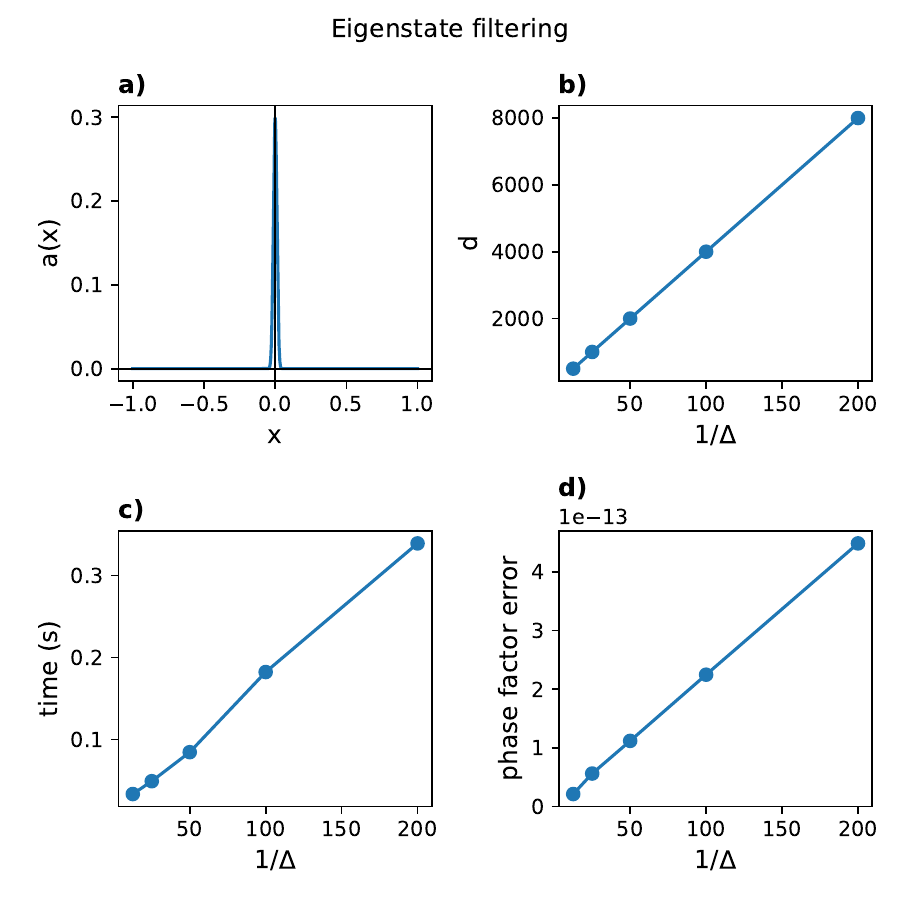}
\caption{Eigenstate filtering. (a) $a(x)$ for $\Delta=0.08$. (b) The degree $d$ of the polynomial approximation as a function of $1/\Delta$. (c) The total phase factor construction time in seconds as a function of $1/\Delta$. (d) The relative $L_\infty$ error of the phase factor reconstruction as a function of $1/\Delta$.}
\label{eig_fil}
\end{figure}

The degree in Figure~\ref{eig_fil} grows exactly linearly because $f$ is itself a polynomial of degree $d = 2k = 40/\Delta$. This is consistent with the $O(\Delta^{-1} \log(1/\epsilon))$ bound of~\cite{lin2020optimal}.

\subsubsection{Matrix Inversion}

We consider the inversion of matrices whose spectrum lies in
\[
D_\kappa=[-1,-1/\kappa]\cup[1/\kappa,1],
\]
where $\kappa$ is the condition number. The QSP problem is to approximate $1/x$ on $D_\kappa$. Following~\cite{ying2022stable}, we use
\[
f(x)=\frac{1-e^{-(5\kappa x)^2}}{x},
\]
whose difference from $1/x$ on $D_\kappa$ is negligible.

We perform experiments with $\kappa=16,64,256,1024$. For $\kappa = 1024$, there are over $50000$ phase factors, matching the largest instance in~\cite{ying2022stable}.

\begin{figure}[!h]
\centering
\includegraphics[width=0.8\linewidth]{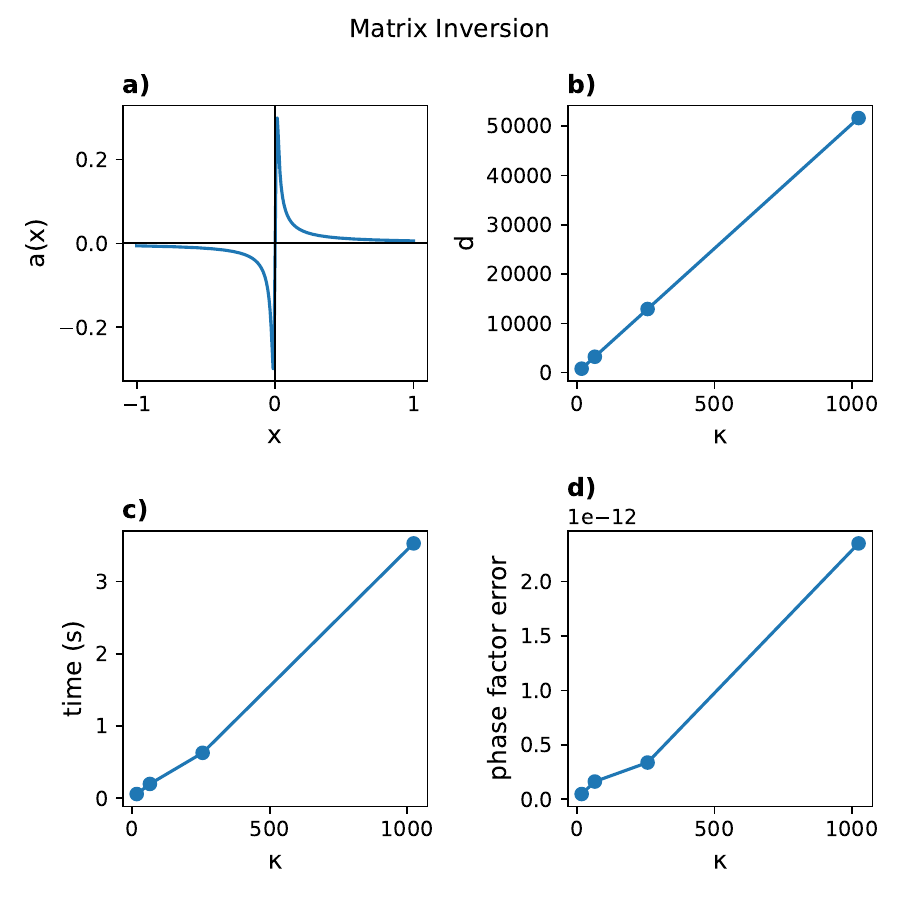}
\caption{Matrix inversion. (a) $a(x)$ for $\kappa=16$. (b) The degree $d$ of the polynomial approximation as a function of $\kappa$. (c) The total phase factor construction time in seconds as a function of $\kappa$. (d) The relative $L_\infty$ error of the phase factor reconstruction as a function of $\kappa$.}
\label{mat_inv}
\end{figure}

The width of the Gaussian smoothing is $1/(5\kappa)$, and the Chebyshev coefficients of a Gaussian decay roughly like $e^{-(k/10\kappa)^2}$~\cite[Chapter 8]{trefethen2019atap}. Hence $d \approx 10 \kappa \sqrt{\ln(1/\epsilon )} \approx 53\kappa$, consistent with the observed ratio $d/\kappa \approx 50$ in Figure~\ref{mat_inv}.

\subsubsection{Fermi--Dirac Operator}

Finally, we consider the Fermi--Dirac function
\[
f(x)=\frac{1-e^{\beta x}}{1+e^{\beta x}}.
\]

We perform experiments with $\beta=100,200,400,800,1600$.

\begin{figure}[!h]
\centering
\includegraphics[width=0.8\linewidth]{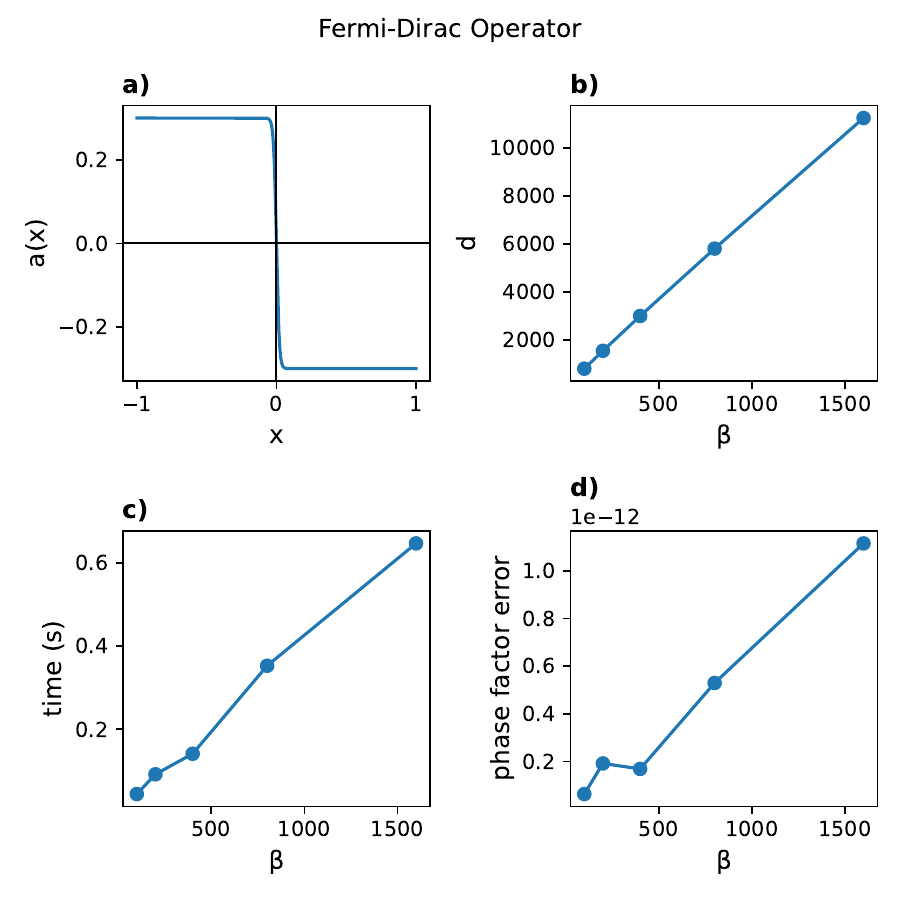}
\caption{Fermi--Dirac operator. (a) $a(x)$ for $\beta=100$. (b) The degree $d$ of the polynomial approximation as a function of $\beta$. (c) The total phase factor construction time in seconds as a function of $\beta$. (d) The relative $L_\infty$ error of the phase factor reconstruction as a function of $\beta$.}
\label{fer_dir}
\end{figure}

The Fermi--Dirac function has poles at distance about $\pi /\beta$ from the real axis, so its Chebyshev coefficients decay like $e^{-\pi k /\beta }$ \cite[Chapter 8]{trefethen2019atap}, where $k$ is the Chebyshev index. Thus $d \approx (\beta/\pi) \log(1/\epsilon)$~\cite{goedecker1999linear}. The predicted slope is roughly $\ln(10^{12})/\pi \approx 8.8$, the same order as the observed slope $d/\beta \approx 7$ in Figure~\ref{fer_dir}.

\section{Conclusion}
\label{sec:conclusion}

\subsection{Comparison to Prony's Method}

Overall, the numerical experiments support the theoretical complexity and the numerical stability of the proposed method. For Hamiltonian simulation and eigenstate filtering, the phase factor errors remain below $2\times10^{-12}$ and $5\times10^{-13}$, respectively, across the tested polynomial degrees. Matrix inversion achieves errors below $3 \times 10^{-12}$ even at degree $51633$, and the Fermi--Dirac examples have errors below $2 \times 10^{-12}$. The function approximation error is below $40\epsilon$, where $\epsilon$ is the chosen threshold. The phase factor error is presumably due to floating-point round-off accumulated in constructing the complementary polynomial and in layer peeling, and it grows slowly with $d$ in all applications. All tested instances complete within a few seconds, and the runtime grows nearly linearly in $d$, as expected from the $O(d\log^2d)$ complexity. Table~\ref{tab:comparison} summarizes the comparison with~\cite{ying2022stable}.

\begin{table}
    \centering
    \begin{tabular}{c|cccc}
          & $e_{\max}$ (Prony) & $e_{\max}$ (Newton) & $t_{\max}$ (Prony) & $t_{\max}$ (Newton) \\ \hline
        HS & $5 \times 10^{-12}$ & $1.4 \times 10^{-12}$ & 20 & 0.3 \\
        EF & $2 \times 10^{-12}$ & $4.5 \times 10^{-13}$ & 30 & 0.3 \\
        MI & $2 \times 10^{-11}$ & $2.4 \times 10^{-12}$ & 300 & 3.5 \\
        FDO &  $4.5 \times 10^{-12}$& $1.1 \times 10^{-12}$ & 70 & 0.6 \\
    \end{tabular}
    \caption{Comparison of Ying's Prony-based algorithm~\cite{ying2022stable} and our Newton--Girard-based algorithm: maximum phase factor error $e_{\max}$ and maximum runtime $t_{\max}$ (in seconds) for Hamiltonian simulation (HS), eigenstate filtering (EF), matrix inversion (MI), and the Fermi--Dirac operator (FDO). The Prony values are taken from~\cite{ying2022stable} rather than from a reimplementation, so they were obtained on different hardware.}
    \label{tab:comparison}
\end{table}

These results indicate that replacing Prony's method with the logarithmic derivative and the Newton--Girard identities preserves high numerical accuracy while yielding a simple and efficient implementation. The decrease in runtime may be largely due to the shift to the divide-and-conquer phase stripping algorithm of~\cite{ni2025inverse}, rather than the replacement of Prony's method.

\subsection{Future Work}

\paragraph{Error analysis of the Newton--Girard step.}
Recovering polynomial coefficients from power sums is notoriously ill-conditioned, since the recursion can significantly amplify errors in $S_k$ when the roots have widely varying magnitudes. Our experiments nonetheless show errors below $3 \times 10^{-12}$ at degrees above $5\times 10^4$, suggesting that the bound $\|a\|_\infty \le 0.3$ keeps the roots of $h$ sufficiently far from the unit circle. A rigorous forward error bound would explain this behavior and identify when it breaks down.

\paragraph{Relaxing the normalization $\|a\|_\infty \le 0.3$.}
Scaling the target polynomial to $\|a\|_\infty = 0.3$ lowers the success probability of the resulting circuit, which then needs more rounds of amplitude amplification. As $\|a\|_\infty \to 1$, the roots of $h$ approach the unit circle, so the logarithmic derivative $l$ becomes nearly singular on the integration contour and its Fourier coefficients decay slowly. Several approaches may help:
\begin{itemize}
    \item choosing the number of samples adaptively rather than fixing it at $40d$;
    \item integrating over deformed contours $|z| = r$ with $r \neq 1$, placed between the inner and outer root sets;
    \item using extended or mixed precision only in the most sensitive stages.
\end{itemize}
Characterizing the achievable accuracy as a function of $1 - \|a\|_\infty$ would make the method more practical.

\paragraph{Generalizations of QSP.}
Our approach sets $b = 0$ and needs only the power sums of the roots inside the unit disk. It therefore applies to any setting that requires a complementary polynomial satisfying $|P|^2 + |Q|^2 = 1$ on the unit circle. Natural next steps are complex-valued targets and generalized QSP with $\mathrm{SU}(2)$ rotations, which lifts the parity constraint and could remove the LCU step used here to combine the even and odd parts.

More broadly, we expect the Kolmogorov (cepstral) approach using logarithmic derivatives to be useful wherever spectral factorization of a positive trigonometric polynomial is required.

\section*{Acknowledgements}

The author thanks Hongkang Ni for providing mentorship and constructive feedback on this paper.

\printbibliography

@article{ying2022stable,
  author  = {Ying, Lexing},
  title   = {Stable factorization for phase factors of quantum signal processing},
  journal = {{Quantum}},
  volume  = {6},
  pages   = {842},
  year    = {2022},
  doi     = {10.22331/q-2022-10-20-842}
}

@article{qspbasics,
  author  = {Wang, Jiasu and Dong, Yulong and Lin, Lin},
  title   = {On the energy landscape of symmetric quantum signal processing},
  journal = {{Quantum}},
  volume  = {6},
  pages   = {850},
  year    = {2022},
  doi     = {10.22331/q-2022-11-03-850}
}

@article{DongLinNiEtAl2024_iqsp,
  author  = {Dong, Yulong and Lin, Lin and Ni, Hongkang and Wang, Jiasu},
  title   = {Infinite quantum signal processing},
  journal = {{Quantum}},
  volume  = {8},
  pages   = {1558},
  year    = {2024},
  doi     = {10.22331/q-2024-12-10-1558}
}

@article{DongLinNiEtAl2024_newton,
  author  = {Dong, Yulong and Lin, Lin and Ni, Hongkang and Wang, Jiasu},
  title   = {Robust Iterative Method for Symmetric Quantum Signal Processing in All Parameter Regimes},
  journal = {SIAM Journal on Scientific Computing},
  volume  = {46},
  number  = {5},
  pages   = {A2951--A2971},
  year    = {2024},
  doi     = {10.1137/23M1598192}
}

@article{alexis2024quantum,
  author  = {Alexis, Michel and Mnatsakanyan, Gevorg and Thiele, Christoph},
  title   = {Quantum signal processing and nonlinear {F}ourier analysis},
  journal = {Revista Matem{\'a}tica Complutense},
  volume  = {37},
  number  = {3},
  pages   = {655--694},
  year    = {2024},
  doi     = {10.1007/s13163-024-00494-5}
}

@article{alexis2024infinite,
  author  = {Alexis, Michel and Lin, Lin and Mnatsakanyan, Gevorg and Thiele, Christoph and Wang, Jiasu},
  title   = {Infinite quantum signal processing for arbitrary {S}zeg{\H o} functions},
  journal = {Communications on Pure and Applied Mathematics},
  volume  = {79},
  year    = {2026},
  doi     = {10.1002/cpa.70007}
}

@misc{ni2024fast,
  author        = {Ni, Hongkang and Ying, Lexing},
  title         = {Fast Phase Factor Finding for Quantum Signal Processing},
  year          = {2024},
  eprint        = {2410.06409},
  archivePrefix = {arXiv}
}

@misc{ni2025inverse,
  author        = {Ni, Hongkang and Sarkar, Rahul and Ying, Lexing and Lin, Lin},
  title         = {Inverse nonlinear fast {F}ourier transform on {SU}(2) with applications to quantum signal processing},
  year          = {2025},
  eprint        = {2505.12615},
  archivePrefix = {arXiv}
}

@misc{lin2025mathematical,
  author        = {Lin, Lin},
  title         = {Mathematical and numerical analysis of quantum signal processing},
  year          = {2025},
  eprint        = {2510.00443},
  archivePrefix = {arXiv},
  note          = {To appear in Proceedings of the ICM 2026}
}

@article{berntson2025complementary,
  author  = {Berntson, Bjorn K. and S{\"u}nderhauf, Christoph},
  title   = {Complementary Polynomials in Quantum Signal Processing},
  journal = {Communications in Mathematical Physics},
  volume  = {406},
  pages   = {161},
  year    = {2025},
  doi     = {10.1007/s00220-025-05302-9}
}

@article{Haah2019,
  author  = {Haah, Jeongwan},
  title   = {Product decomposition of periodic functions in quantum signal processing},
  journal = {{Quantum}},
  volume  = {3},
  pages   = {190},
  year    = {2019},
  doi     = {10.22331/q-2019-10-07-190}
}

@inproceedings{gilyen2019quantum,
  author    = {Gily{\'e}n, Andr{\'a}s and Su, Yuan and Low, Guang Hao and Wiebe, Nathan},
  title     = {Quantum singular value transformation and beyond: Exponential improvements for quantum matrix arithmetics},
  booktitle = {Proceedings of the 51st Annual ACM SIGACT Symposium on Theory of Computing},
  pages     = {193--204},
  year      = {2019},
  doi       = {10.1145/3313276.3316366}
}

@article{dong2021efficient,
  author  = {Dong, Yulong and Meng, Xiang and Whaley, K. Birgitta and Lin, Lin},
  title   = {Efficient phase-factor evaluation in quantum signal processing},
  journal = {Physical Review A},
  volume  = {103},
  number  = {4},
  pages   = {042419},
  year    = {2021},
  doi     = {10.1103/PhysRevA.103.042419}
}

@article{low2017optimal,
  author  = {Low, Guang Hao and Chuang, Isaac L.},
  title   = {Optimal {H}amiltonian Simulation by Quantum Signal Processing},
  journal = {Physical Review Letters},
  volume  = {118},
  number  = {1},
  pages   = {010501},
  year    = {2017},
  doi     = {10.1103/PhysRevLett.118.010501}
}

@article{lin2020optimal,
  author  = {Lin, Lin and Tong, Yu},
  title   = {Optimal polynomial based quantum eigenstate filtering with application to solving quantum linear systems},
  journal = {{Quantum}},
  volume  = {4},
  pages   = {361},
  year    = {2020},
  doi     = {10.22331/q-2020-11-11-361}
}

@article{childs2017quantum,
  author  = {Childs, Andrew M. and Kothari, Robin and Somma, Rolando D.},
  title   = {Quantum Algorithm for Systems of Linear Equations with Exponentially Improved Dependence on Precision},
  journal = {SIAM Journal on Computing},
  volume  = {46},
  number  = {6},
  pages   = {1920--1950},
  year    = {2017},
  doi     = {10.1137/16M1087072}
}

@article{goedecker1999linear,
  author  = {Goedecker, Stefan},
  title   = {Linear scaling electronic structure methods},
  journal = {Reviews of Modern Physics},
  volume  = {71},
  number  = {4},
  pages   = {1085--1123},
  year    = {1999},
  doi     = {10.1103/RevModPhys.71.1085}
}

@article{brent1978fast,
  author  = {Brent, Richard P. and Kung, Hsiang T.},
  title   = {Fast Algorithms for Manipulating Formal Power Series},
  journal = {Journal of the ACM},
  volume  = {25},
  number  = {4},
  pages   = {581--595},
  year    = {1978},
  doi     = {10.1145/322092.322099}
}

@book{trefethen2019atap,
  author    = {Trefethen, Lloyd N.},
  title     = {Approximation Theory and Approximation Practice, Extended Edition},
  publisher = {Society for Industrial and Applied Mathematics},
  address   = {Philadelphia, PA},
  year      = {2019},
  isbn      = {978-1-611975-93-2},
  doi       = {10.1137/1.9781611975949}
}

\end{document}